\documentclass[final,3p,times,onecolumn]{elsarticle}
\usepackage{times}
\usepackage{amsmath}
\usepackage{graphicx}
\newcommand{\ptl}{\partial}
\journal{Journal of Franklin Institute}

\begin{document}

\begin{frontmatter}



\title{Channel Equalization and Beamforming for Quaternion-Valued Wireless Communication Systems}


%\author[bit,shef]{Xirui Zhang}
%\ead{xrzhang@bit.edu.cn}
\author[shef]{Wei Liu\corref{cor1}}
\cortext[cor1]{Corresponding author: Tel.: +44-114-2225813; fax: +44-114-2225834.}
\ead{w.liu@sheffield.ac.uk}
%\author[bit]{Yougen Xu}
%\ead{yougenxu@bit.edu.cn}
%\author[bit]{Zhiwen Liu}
%\ead{zwliu@bit.edu.cn}
%%\author[shef]{Wei Liu}
%\address[bit]{School of Information and Electronics, Beijing Institute of Technology\\ Beijing, 100081, China}
\address[shef]{Communication Research Group, Department of Electronic and Electrical Engineering, University of Sheffield\\
Sheffield, S1 3JD, United Kingdom.}
%-----Abstract-----%
\begin{abstract}
Quaternion-valued wireless communication systems have been studied in the past. Although progress has been made in this promising area, a crucial missing link  is lack of effective and efficient quaternion-valued signal processing algorithms for channel equalization and beamforming. With most recent developments in quaternion-valued signal processing, in this work, we fill the gap to solve the  problem by studying two quaternion-valued adaptive algorithms: one is the reference signal based quaternion-valued least mean square (QLMS) algorithm and the other one is the quaternion-valued constant modulus algorithm (QCMA). The quaternion-valued Wiener solution for possible block-based calculation is also derived. Simulation results are provided to show the working  of the system.
\end{abstract}

\end{frontmatter}

%%%%%%%%%%%%%%%%%%%%%%%%%%%%%%%%%%%%%%%%%%%%%%%%%%%%%%%%%%%%%%%%%%%%%%%%%%%%%%%%%%%%%
\section{Introduction}
\label{sec:introdcution}

Increasing the capacity of a wireless communication system has always been a focus of the wireless communications research community. It is well-known that polarisation diversity can be exploited to mitigate the multipath effect to maintain a reliable communication link with an acceptable quality of service (QoS), where a pair of antennas with orthogonal polaristion directions is employed at both the transmitter and the receiver sides. However, the traditional diversity scheme aims to achieve a single reliable channel link between the transmitter and the receiver, while the same information is transmitted at the same frequency but with different polarisations, i.e. two channels. This is not an effective use of the precious spectrum resources as the two channels could be used to transmit different data streams simultaneously. For example, we can design a four-dimensional (4-D) modulation scheme across the two polarisation diversity channels using a quaternion-valued representation, as proposed in \cite{isaeva95a}. An earlier version of quaternion-valued 4-D modulation scheme based on two different frequencies was proposed in \cite{zetterberg77a}. However, due to the change of polarisation of the transmitted radio frequency signals during the complicated propagation process including multipath, reflection, refraction, etc, interference will be caused to each other at the two differently polarised receiving antennas. To solve the problem, efficient signal processing methods and algorithms for channel equalization and interference suppression/beamforming are needed for practical implementation of the proposed 4-D modulation scheme.

Recently, quaternion-valued signal processing has been introduced and studied in details to solve  problems related to three or four-dimensional signals \cite{bihan04a}, such as vector-sensor array signal processing~\cite{miron06a,gong11a,tao13a,liu14e,liu14j,liu15a}, and wind profile prediction~\cite{liu14f}. With most recent developments in this area, especially the derivation of quaternion-valued gradient operators and the quaternion-valued least mean square (QLMS) algorithm \cite{liu14f,liu14h,liu14p}, we are now ready to effectively solve the 4-D equalisation and interference suppression/beamforming problem associated with the proposed  4-D modulation scheme. Now the dual-channel effect on the transmitted signal can be modeled by a quaternion-valued infinite impulse response (IIR) or finite impulse response (FIR) filter. At the receiver side, for channel equalisation, we can employ a quaternion-valued adaptive algorithm to recover the original 4-D signal, which inherently also performs an interference suppression operation to separate the original two 2-D signals. Moreover, multiple antenna pairs can be employed at the receiver side to perform the traditional beamforming task to suppress other interfering signals.

In particular, two representative quaternion-valued equalisation/beamforming algorithms will be derived: the first one is the quaternion-valued Wiener filter as a follow-up to the previously derived  QLMS algorithm for reference signal based equalisation/beamforming, and the second one is the quaternion-valued constant modulus algorithm (QCMA) for blind equalisation/beamforming. Compared to the summary contribution in \cite{liu14n}, in addition to the detailed analytical modeling steps, the main difference is the GCMA algorithm and the related simulations. Although quaternion-valued wireless communication employing multiple antennas has been studied before, such as the design of orthogonal space-time-polarization block code in \cite{wysocki09a}, to our best knowledge, it is the first time to study the quaternion-valued equalization and interference suppression/beamforming problem in this context. Moreover, the dual-polarised antenna pair or an array of them has a similar structure to the well-studied vector sensors or sensor arrays \cite{compton81b,nehorai99a,zoltowski00a,liu14l}, where they are used mainly for traditional array signal processing applications. Although the recently developed quaternion-valued array signal processing algorithms based on such traditional array applications employed a quaternion-valued array model~\cite{miron06a,gong11a,tao13a,liu14e}, the desired signals are still traditional complex-valued signals, instead of  quaternion-valued communication signals.

In the following, the 4-D modulation scheme based on two orthogonally polarised antennas will be introduced in Sec. \ref{sec:modulation} and the required quaternion-valued equalisation and inter-channel interference suppression solution and their extension to multiple dual-polarised antennas are presented in Sec. \ref{sec:equal_beam}. Simulation results are provided in Sec. \ref{sec:sim}, followed by conclusions in Sec. \ref{sec:concl}.

%%%%%%%%%%%%%%%%%%%%%%%%%%%%%%%%%%%%%%%%%%%%%%%%%%%%%%%%%%%%%%%%%%%%%%%%%%%%%%%%%%%%%%%%%%%%%%%

\section{Quaternion-Valued 4-D Modulation}
\label{sec:modulation}
%\subsection{Four-dimensional modulation}
%\label{sec:proposed_modulation}

\begin{figure}
	\centering
	\includegraphics[scale=0.72]{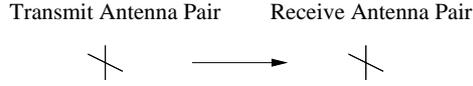}
	\caption{Wireless communication employing polarisation diversity, where both the transmitter and the receiver sides are equipped with a pair of antennas with different polarisation directions.}
	\label{fig:TRantennapair}
\end{figure}

In traditional polarisation diversity scheme, as shown in Fig.~\ref{fig:TRantennapair},
each side is equipped with two antennas with orthogonal polarisation directions and the signal being transmitted is two-dimensional, i.e. complex-valued with one real part and one imaginary part. In the quaternion-valued modulation scheme, the signal is modulated across the two antennas to generate a 4-D modulated signal. Such a signal can be conveniently represented mathematically by a quaternion \cite{hamilton66a,kantor89a}.
A quaternion is a hypercomplex number defined as
\begin{equation}
\label{eq:quaternion}
    q=q_0 + iq_1 + jq_2 + kq_3\;,
\end{equation}
where $q_0$ is the real part of the quaternion, and $q_1$, $q_2$ and
$q_3$ are the three imaginary components with their corresponding imaginary units $i$, $j$ and $k$, respectively. The conjugate of a quaternion, denoted by $q^{*}$, is defined as
\begin{equation}
q^{*}= q_0-iq_1-jq_2-kq_3\;.
\end{equation}
$i$, $j$ and $k$ satisfy the following conditions
\begin{eqnarray}
  &&ii=jj=kk=-1\;,\\
  &&ij=-ji=k\;, jk=-kj=i\;, ki=-ik=j\;.
\end{eqnarray}
%%Finally $\{.\}^{\triangleleft}$ denotes the conjugate transpose ofquaternionic vectors and matrices.
As a result, quaternionic multiplications are noncommutative.

As an example, corresponding to the 4-QAM (Quadrature Amplitude Modulation) in the two-dimensional case, for the 4-D modulation scheme, $q_0$, $q_1$, $q_2$ and $q_3$ can take values of either $1$ or $-1$, representing $16$ different symbols. We can call this scheme 16-QQAM (Quaternion-valued QAM) or 16-$\mbox{Q}^2\mbox{AM}$.

\section{Quaternion-Valued Equalization and Interference Suppression/Beamforming}
\label{sec:equal_beam}
\subsection{Channel model}
\begin{figure}
	\centering
	\includegraphics[scale=0.49]{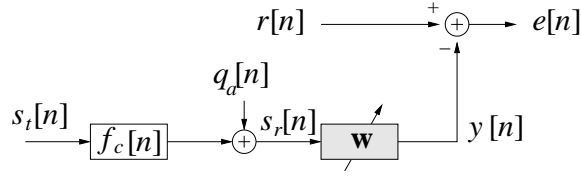}
	\caption{Channel model and the reference signal based equalizer for quaternion-valued signals.}
	\label{fig:channelmodel}
\end{figure}

The signal transmitted by the two antennas will go through the channel with all kinds of effects and arrive at the receiver side, where the two antennas with orthogonal polarisation directions (Note that orthogonal polarisation may not give the best performance for a specific scenario) will pick up the two signals. Again the four components of the received signal can be represented by another quaternion. We use $s_t[n]$ and $s_r[n]$ to represent the transmitted and received 4-D quaternion-valued signals, respectively. Then the channel effect can be modeled by a filter with quaternion-valued impulse response $f_c[n]$, i.e.
\begin{equation}
\label{eq:channelmodel}
    s_r[n]=s_t[n]*f_{c}[n]+q_a[n]\;,
\end{equation}
where $q_a[n]$ is the quaternion-valued additive noise, as shown in Fig.~\ref{fig:channelmodel}.

%%%%%%%%%%%%%%%%%%%%%%%%%%%%%%%%%%%%%%%%%%%%%%%%%%%%%
\subsection{Reference signal based approach}
To recover $s_t[n]$ from $s_r[n]$ or estimate the channel, as in the 2-D case (complex-valued), we can design a quaternion-valued equalizer. One choice is a reference signal based equalizer, among many others corresponding to the complex-valued case. Now assume we have a reference signal $r[n]$ available. Then we can employ the standard adaptive filtering structure shown in the second half of Fig.~\ref{fig:channelmodel} and update the equalizer coefficient vector $\mathbf{w}$ with a length of $L$ by minimising the mean square value of the error signal $e[n]$~\cite{liu14f,liu14h,liu14p}.

The cost function is given by
\begin{equation}
J[n]=e[n]e^*[n]\;,
\end{equation}
where
\begin{equation}
e[n]=r[n]-y[n]=r[n]-\textbf{w}^{T}\textbf{s}_r[n]\;.
\end{equation}
with $\textbf{w}$ being the equalizer coefficient vector and $\textbf{s}_r[n]$ holding the corresponding received signal samples from $s_r[n]$
\begin{eqnarray}
\textbf{w}&=&[w_0, w_1, \cdots, w_{L-1}]^T\nonumber\\
\textbf{s}_r[n]&=&[s_r[n], s_r[n-1], \cdots, s_r[n-L+1]]^{T}\;.
\end{eqnarray}

For a general quaternion-valued function $f(q)$ of a quaternion $q$, the gradient of $f$ with respect to $q$ is defined as \cite{liu14f,liu14h,liu14p}
\begin{align}
  \frac{\ptl f}{\ptl q}&:=\frac{1}{4}\left(\displaystyle\frac{\partial{f}}{\partial q_0}-
\frac{\partial{f}}{\partial q_1}i -\displaystyle\frac{\partial{f}}{\partial q_2}j -
\displaystyle\frac{\partial{f}}{\partial q_3}k \right), \label{eq:rdfdq} \\
  \frac{\ptl f}{\ptl q^*}&:=\frac{1}{4}\left(\displaystyle\frac{\partial{f}}{\partial q_0}+
\frac{\partial{f}}{\partial q_1}i +
\displaystyle\frac{\partial{f}}{\partial q_2}j +
k\displaystyle\frac{\partial{f}}{\partial q_3}k \right), \label{eq:rdfdq1}
\end{align}

Following the derivations in \cite{liu14f,liu14h,liu14p}, we have the gradient of $J[n]$ with respect to the coefficient vector as follows
\begin{equation}
\nabla_{\textbf{w}}J[n]=-\frac{1}{2}\textbf{s}_r[n]e^*[n],
\label{eqn:gradient}
\end{equation}
which leads to the following update equation for the coefficient vector with a step size of $\mu$, i.e. the QLMS algorithm:
\begin{eqnarray}
\textbf{w}[n+1] &=& \textbf{w}[n]-\mu (\nabla_{\textbf{w}}J[n])^{*}\\
 &=&\textbf{w}[n]+\mu(e[n]\textbf{s}_r^{*}[n])\;.
\label{eq:QLMSupdate}
\end{eqnarray}

For a solution equivalent to the classic Wiener filter in the complex-valued case, based on the instantaneous gradient result of \eqref{eqn:gradient}, the optimum solution $\textbf{w}_{opt}$ should satisfy
\begin{eqnarray}
E\{-\frac{1}{2}\textbf{s}_r[n]e^*[n]\}=0,
\label{eqn:gradient1}
\end{eqnarray}
i.e.
\begin{eqnarray}
E\{\textbf{s}_r[n]e^*[n]\}&=&E\{\textbf{s}_r[n]r^*[n]-\textbf{s}_{r}[n]\textbf{s}^H_r[n]\textbf{w}^*_{opt}\}\nonumber\\
&=&\textbf{p}-\textbf{R}_{s_r}\textbf{w}_{opt}^*=0\;,
\label{eqn:gradient2}
\end{eqnarray}
where the cross-correlation vector $\textbf{p}=E\{\textbf{s}_r[n]r^*[n]\}$ and the covariance matrix  $\textbf{R}_{s_r}=E\{\textbf{s}_r[n]\textbf{s}^H_r[n]\}$.
Then we have
\begin{equation}
\textbf{R}_{s_r}\textbf{w}_{opt}^*=\textbf{p}\;\;\Rightarrow\;\; \textbf{w}^*_{opt}=\textbf{R}_{s_r}^{-1}\textbf{p}\;.
\label{eqn:wiener}
\end{equation}
We can use the above equation to obtain the optimum weight vector directly.

Now consider the multiplication of two quaternions $a$ and $b$ with $c=ab$. They are expressed as
\begin{eqnarray}
\label{eqn:quaternionmultiplication}
    a&=&a_0 + ia_1 + ja_2 + ka_3\nonumber\\
    b&=&b_0 + ib_1 + jb_2 + kb_3\nonumber\\
    c&=&c_0 + ic_1 + jc_2 + kc_3
\end{eqnarray}
where the subscripts indicate the corresponding components of the quaternion.
Then we have
\begin{eqnarray}
\label{eqn:quaternionmultiplication1}
    c_0&=&b_0a_0-b_1a_1-b_2a_2-b_3a_3\nonumber\\
    c_1&=&b_0a_1+b_1a_0-b_2a_3+b_3a_2\nonumber\\
    c_2&=&b_0a_2+b_1a_3+b_2a_0-b_3a_1\nonumber\\
    c_3&=&b_0a_3-b_1a_2+b_2a_1+b_3a_0
\end{eqnarray}
Using this result, we can obtain the solution to \eqref{eqn:wiener} using real-valued matrix operations. First we define the following two vectors
 \begin{eqnarray}\label{eqn:w_hat}
  \hat{\textbf{w}}_{opt} &=& [R(w^\ast_{opt,0}), -I(w^\ast_{opt,0}), -J(w^\ast_{opt,0}), -K(w^\ast_{opt,0}),\nonumber\\
  &&\cdots, R(w^\ast_{opt,L-1}), -I(w^\ast_{opt,L-1}), -J(w^\ast_{opt,L-1}), -K(w^\ast_{opt,L-1})]|^{T}\nonumber\\
    \hat{\textbf{p}}&=&[R(\textbf{p})^T, I(\textbf{p})^T, J(\textbf{p})^T,  K(\textbf{p})^T]^T\;,
\end{eqnarray}
where $w_{opt,l}$, $l=0, \cdots, L-1$ is the $l$-th element of the optimum weight vector $\textbf{w}_{opt}$, $R(\cdot)$, $I(\cdot)$, $J(\cdot)$, and $K(\cdot)$ are the operation of taking the real and three imaginary components of the quaternion inside the brackets, respectively.
We also define
\begin{equation}\label{eqn:RSR}
    \hat{\textbf{R}}_{s_r} = [\hat{\textbf{R}}_{s_r,0}; \cdots;   \hat{\textbf{R}}_{s_r,L-1}]\;,
\end{equation}
with
\begin{equation}\label{eqn:RSRL}
    \hat{\textbf{R}}_{s_r,l}=\left(
                       \begin{array}{cccc}
                         R(\textbf{r}_{l}) & -I(\textbf{r}_{l})& -J(\textbf{r}_{l})& -K(\textbf{r}_{l})\\
                         I(\textbf{r}_{l}) & R(\textbf{r}_{l})& -K(\textbf{r}_{l})& +J(\textbf{r}_{l})\\
                         J(\textbf{r}_{l}) & K(\textbf{r}_{l})& R(\textbf{r}_{l})& -I(\textbf{r}_{l})\\
                         K(\textbf{r}_{l}) & -J(\textbf{r}_{l})& I(\textbf{r}_{l})& R(\textbf{r}_{l})
                       \end{array}
                     \right)
\end{equation}
for $l=0, 1, \cdots, L-1$, where $\textbf{r}_{l}$ is the $l$-th column vector of the  covariance matrix $\textbf{R}_{s_r}$.

Then, according to \eqref{eqn:wiener}, we have the following relationship
\begin{equation}\label{eqn:realwiener}
    \hat{\textbf{R}}_{s_r}\hat{\textbf{w}}_{opt} = \hat{\textbf{p}}\;,
\end{equation}
where all the matrix and vectors involved are real-valued. Then $\hat{\textbf{w}}_{opt}$ is obtained by
\begin{equation}\label{eqn:realwiener1}
    \hat{\textbf{w}}_{opt} = \hat{\textbf{R}}_{s_r}^{-1}\hat{\textbf{p}}\;.
\end{equation}
From $\hat{\textbf{w}}_{opt}$, we can then easily deduce $\textbf{w}_{opt}$.

%%%%%%%%%%%%%%%%%%%%%%%%%%%%%%%%%%%%%%%%%%%%%%
\subsection{Constant modulus based approach}
When a reference signal is not available, it is still possible to perform equalisation and interference suppression/beamforming by employing other properties of the signals. An algorithm designed to work without knowledge of the transmitted signals falls into the category of blind equalisation and beamforming approaches \cite{ding01a,vantrees02a,liu10g}. One representative blind equalisation algorithm in traditional communication systems is the constant modulus algorithm (CMA)~\cite{gooch86a,krim96,godara97b,johnson98a,chen07d,liu10k}.

There are many variations to this algorithm and the basic form is based on minimizing the following cost function
\begin{equation}\label{eq:cma1}
J_{CM}=\frac{1}{4}E\{(|y[n]|^2-\gamma)^2\}\;,
\end{equation}
where $y[n]$ is the output of the equalizer and $\gamma$ is the dispersion constant, defined by  $\gamma=\frac{E\{s_k\}^4}{E\{s_k\}^2}$ with $s_k$ being symbols of the modulation scheme.

For our quaternion-valued 4-D wireless communication system, we can develop a similar quaternion-valued CMA (QCMA) for blind equalization and beamforming. Taking the gradient of $J_{CM}$ with respect to the quaternion-valued coefficient vector $\textbf{w}$, and using the third chain rule of the restricted HR gradient operation given in \cite{liu14p} as the intermediate function is real-valued, we have
\begin{eqnarray}
\nabla_{\textbf{w}}J_{CM}&=&\frac{1}{2}E\{(y[n]\cdot y^\ast[n]-\gamma) \frac{\ptl (y[n]\cdot y^\ast[n]-\gamma)}{\ptl \textbf{w}}\}\nonumber\\
&=&\frac{1}{2}E\{(y[n]\cdot y^\ast[n]-\gamma) \frac{\ptl (y[n]\cdot y^\ast[n])}{\ptl \textbf{w}}\}\nonumber\\
&=&\frac{1}{2}E\{(y[n]\cdot y^\ast[n]-\gamma) \frac{\ptl ({\textbf{w}^{T}} {\textbf{s}_r[n]}{\textbf{s}_r^{H}[n]}{\textbf{w}^{*}})}{\ptl \textbf{w}}\}\nonumber\\
\label{eqn:gradientCM0}
\end{eqnarray}
Using the result of \cite{liu14f}, we have
\begin{equation}
\frac{\partial({\textbf{w}^{T}} {\textbf{s}_r[n]}{\textbf{s}_r^{H}[n]}{\textbf{w}^{*}})}{\partial {\textbf{w}}}=\frac{1}{2}{\textbf{s}_r[n]}{\textbf{y}^\ast[n]}
\label{eq:partresult}
\end{equation}
Then we have
\begin{equation}
\nabla_{\textbf{w}}J_{CM}=\frac{1}{4}E\{(y[n]\cdot y^\ast[n]-\gamma) {\textbf{s}_r[n]}{\textbf{y}^\ast[n]}\}
\label{eqn:gradientCM1}
\end{equation}
Using the instantaneous gradient $\hat{\nabla}_{\textbf{w}}J_{CM}$ to replace $\nabla_{\textbf{w}}J_{CM}$, we then obtain the final update equation for our quaternion-valued constant modulus algorithm
\begin{eqnarray}
\textbf{w}[n+1] &=& \textbf{w}[n]-\mu (\nabla_{\textbf{w}}J_{CM})^{*}\\
 &=&\textbf{w}[n]- \mu(|y[n]|^2-\gamma) y[n]\textbf{s}_r^\ast[n]\;,
\label{eq:update_weight_vector}
\end{eqnarray}
where the constant $\frac{1}{4}$ has been absorbed into the step size $\mu$.

%%%%%%%%%%%%%%%%%%%%%%%%%%%%%%%%%

\subsection{Extension to multiple antennas and MIMO systems}
\begin{figure}
	\centering
	\includegraphics[scale=0.69]{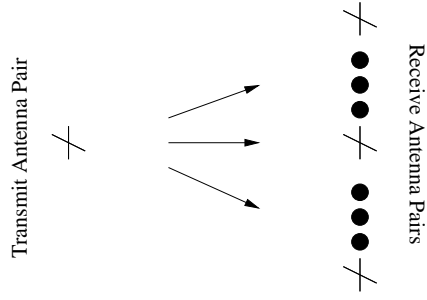}
	\caption{Multiple antenna pairs at the receiver side, where each pair is composed of two differently polarised antennas.}
	\label{fig:pairarray}
\end{figure}

\begin{figure}
	\centering
	\includegraphics[scale=0.69]{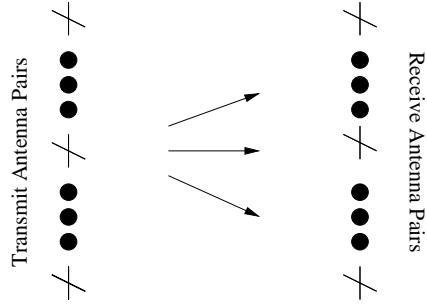}
	\caption{A MIMO system with multiple antenna pairs at both the transmitter and the receiver sides, where each pair is composed of two differently polarised antennas.}
	\label{fig:MIMOarray}
\end{figure}

\begin{figure}
	\centering
	\includegraphics[scale=0.5]{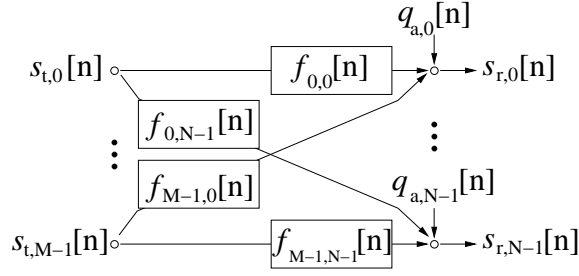}
	\caption{Channel model for a quaternion-valued MIMO system.}
	\label{fig:MIMOchannelmodel}
\end{figure}

We can extend this design to multiple antenna pairs for beamforming to suppress other quaternion-valued interfering signals,  as shown in Fig.~\ref{fig:pairarray}, or Fig.~\ref{fig:MIMOarray} for a general multiple-input-multiple-output (MIMO) system. The channel model for an $M\times N$ system is shown in Fig.~\ref{fig:MIMOchannelmodel}, where $s_{t,m}[n]$, $m=0, 1, \cdots, M-1$ is the transmitted signal, while $s_{r,l}[n]$, $l=0, 1, \cdots, N-1$ is the received signal. $q_{a,l}[n]$, $l=0, 1, \cdots, N-1$ is the added channel noise and $f_{m,l}[n]$ is the quaternion-valued channel impulse response between the $m$-th transmit antenna pair and the $l$-th receive antenna pair.

Mathematically, we have
\begin{equation}
     \mathbf{s}_r[n] = \mathbf{A}^T\cdot\mathbf{s}_t[n]+\mathbf{q}_a[n]\;,
\label{eqn:mimo_channelmodel}
\end{equation}
where
\begin{eqnarray}
     \mathbf{A}&=& \left[ \begin{array}{cccc} f_{0,0}[n] & f_{0,1}[n] & \ldots & f_{0,N-1}(t)\\
      f_{1,0}[n] & f_{1,1}[n] & \ldots & f_{1,N-1}(t)\\
      \vdots & \vdots & \ddots & \vdots \\
       f_{M-1,0}[n] & f_{M-1,1}[n] & \ldots & f_{M-1,N-1}[n]
      \end{array} \right]\;,\nonumber\\
      \textbf{s}_t[n]&=&[s_{t,0}[n], s_{t,1}[n], \cdots, s_{t,M-1}]^T\nonumber\\
      \textbf{s}_r[n]&=&[s_{r,0}[n], s_{r,1}[n], \cdots, s_{r,N-1}]^T\nonumber\\
      \textbf{q}_a[n]&=&[q_{a,0}[n], q_{a,1}[n], \cdots, q_{a,N-1}]^T
      \label{eqn:mimo_channelmatrix}
\end{eqnarray}

Applying an $L\times 1$ weight vector $\mathbf{w}_{m,l}=[w_{m,l,0}, w_{m,l,1}, \cdots, w_{m,l,L-1}]^T$, $m=0, 1, \cdots, M-1$, $l=0, 1, \cdots, N-1$, to the received signal $s_{r,l}[n]$ and then combining the $N$ corresponding outputs together, we obtain one of the estimated signals $y_m[n]$, i.e.
\begin{equation}
y_m[n]=\textbf{w}_m^{T}\mathbf{\hat{s}}_r[n]\;,
\end{equation}
where
\begin{eqnarray}
\mathbf{w}_m&=&[\mathbf{w}_{m,0}^T, \mathbf{w}_{m,1}^T, \cdots, \mathbf{w}_{m,N-1}^T]^T\nonumber\\
\mathbf{\hat{s}}_r[n]&=&[\mathbf{s}_{r,0}^T[n], \mathbf{s}_{r,1}^T[n], \cdots, \mathbf{s}_{r,N-1}^T[n]]^T\nonumber\\
\mathbf{s}_{r,l}[n]&=&[s_{r,l}[n], s_{r,l}[n-1], \cdots, s_{r,,l}[n-L+1]]^T\;.
\end{eqnarray}

The optimum weight vector for $\mathbf{w}_m$ can be obtained using either the QLMS algorithm or the QCMA introduced before so that $y_m[n]$ becomes a good estimate of one of the transmitted  signals.

%%%%%%%%%%%%%%%%%%%%%%%%%%%%%%%%%%%%%%%%%%%%%%%%%
\section{Simulation Results}
\label{sec:sim}
In the following, we  give three sets of simulation results. The first two are for the QLMS algorithm and the third one for the QCMA.
\begin{figure}
	\centering
	\includegraphics[scale=0.56]{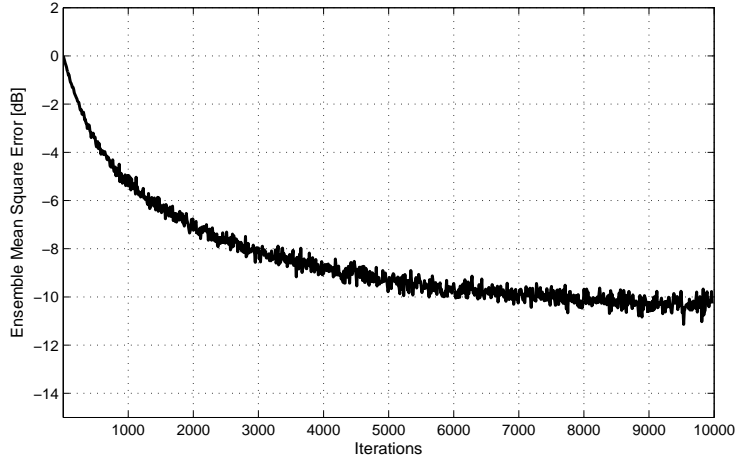}
	\caption{Learning curve of the quaternion-valued equalizer using the QLMS algorithm.}
	\label{fig:learningcurve}
\end{figure}

The first set of simulations is based on the structure in Fig.~\ref{fig:channelmodel} and it is a reference-signal based channel equalization problem. The signal transmitted is 16-$\mbox{Q}^2\mbox{AM}$ modulated and the SNR at the receiver side is 20 dB with quaternion-valued Gaussian noise. The channel impulse response $f_c[n]$ is a 4-tap quaternion-valued FIR filter with a Gaussian-distributed coefficients value, which is generated randomly for each run. The equalizer filter has a length of $15$. The learning curve based on averaging $200$ simulation runs using the QLMS algorithm is shown in Fig.~\ref{fig:learningcurve} with a step size $\mu=0.000012$, with about $-10$ dB error at the steady state, indicating a reasonable channel estimation result. Note that the optimum $\mu$ for each channel realization is different and a fixed $\mu$ for all randomly generated channel impulse responses will lead to a much less favorable result, which is why the steady state error is relatively large. We have also calculated the corresponding bit error rate (BER) based on this non-optimum $\mu$ and it is about $0.29\%$.

\begin{figure}
	\centering
	\includegraphics[scale=0.6]{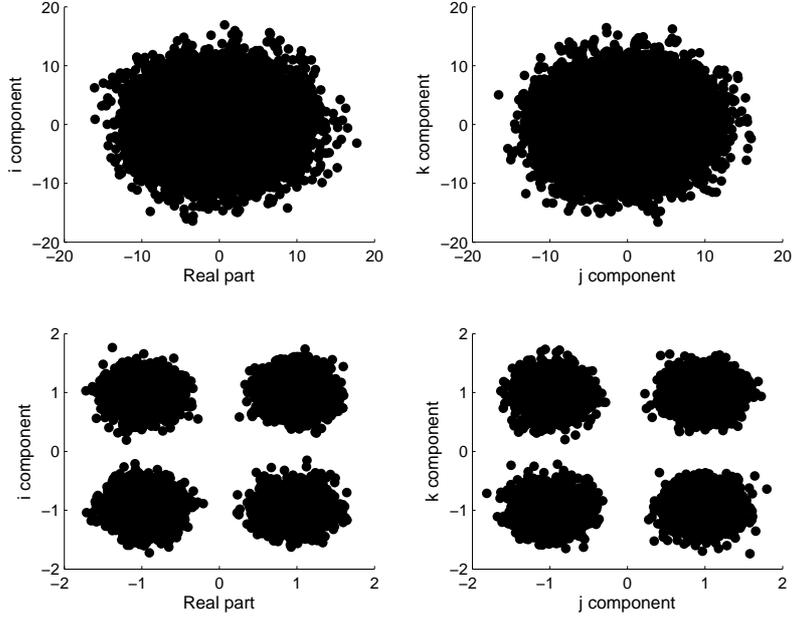}
	\caption{Scatter plots before and after equalization using the QLMS algorithm.}
	\label{fig:Qequal_Scatter}
\end{figure}

However, if we focus on one specific channel response, such as the one given below
\begin{equation}
  \left[ \begin{array}{cccc}
   -1.1696 &   0.4989 &   1.0571 &   0.4911\\
   -0.1135 &  -0.3639 &  -0.3791 &   0.8653\\
    0.0958 &  -0.2515  &  1.4957   & 0.1552\\
   -0.9583 &  -0.6047 &   0.5183 &  -0.8966
 \end{array} \right]\;,
  \label{eqn:sim_fc}
\end{equation}
where each column gives the four components of the quaternion-valued coefficient of $f_c[n]$, and use the same value of $\mu=0.000012$, we will be able to achieve a zero BER result, as shown by the scatter plots of constellation before and after equalization in Fig.~\ref{fig:Qequal_Scatter}. Since we can not show the 4-D scatter plot directly, we have split it into two 2-D plots in Fig.~\ref{fig:Qequal_Scatter}, where the upper row is for the scatter plots before equalization, and the lower row for the plots after equalization. We can clearly see that the equalization operation has been successful.

\begin{figure}
	\centering
	\includegraphics[scale=0.56]{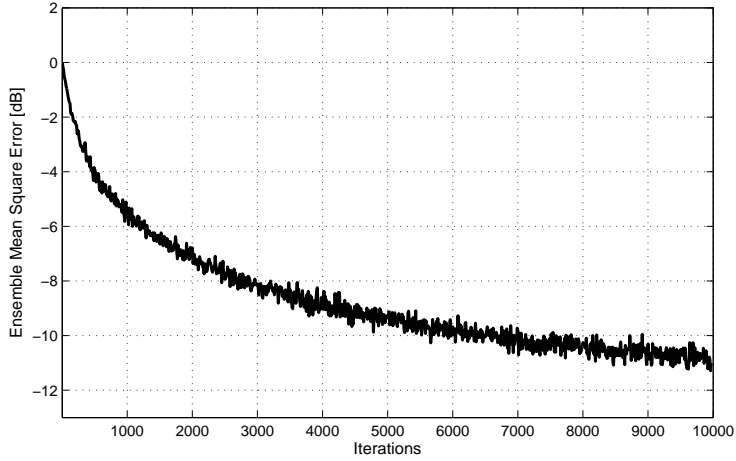}
	\caption{Learning curve of the quaternion-valued equalizer/MIMO beamformer using the QLMS algorithm.}
	\label{fig:learningMIMO}
\end{figure}
In the second set of simulations, we consider a $2\times2$ MIMO array and the two transmitted quaternion-valued signals have the same normalized power, with an SNR of 20 dB at the receiver side. All the other parameters are the same as the first one, except that now the step size has been changed to $\mu=0.0000006$. Fig.~\ref{fig:learningMIMO} shows the result, again a reasonable performance with such a fixed non-optimum step size.

Moreover, we have used the MATLAB function `tic' and `toc' to calculate the running time for the two scenarios. Based on a Windows 7 computer with Intel Core i5-2467M CPU (1.60GHz) and 4GB memory, it took about 4.9 seconds for each run of the first scenario and 10.6 seconds for each run of the second scenario. This time is dependent on the stepsize adopted for the algorithm. A larger stepsize will significantly reduce the time required for convergence. The reason for our current step size value (which is quite small) is to make sure the algorithm will converge for all channel realizations in the simulation. In practice, this step size could be normalized according to the received signal powers to give a much faster convergence rate, in a similar way to the normalized LMS algorithm in literature for complex-valued signals~\cite{haykin96a,liu10g}.

Now we consider an example for the QCMA. A well-known problem with constant modulus based algorithm is that it may converge to local minima depending on its initial values. We find that it is very difficult to initialize the algorithm properly to perform both the equalization and beamforming tasks simultaneously based on the general MIMO structure. So here we only show a case for blind equalization using the QCMA. The setting is the same as the first set of simulations. Since we are using the 16-$\mbox{Q}^2\mbox{AM}$ modulation scheme, we have chosen $\gamma=4$. The stepsize is $\mu=0.0012$.

\begin{figure}
	\centering
	\includegraphics[scale=0.56]{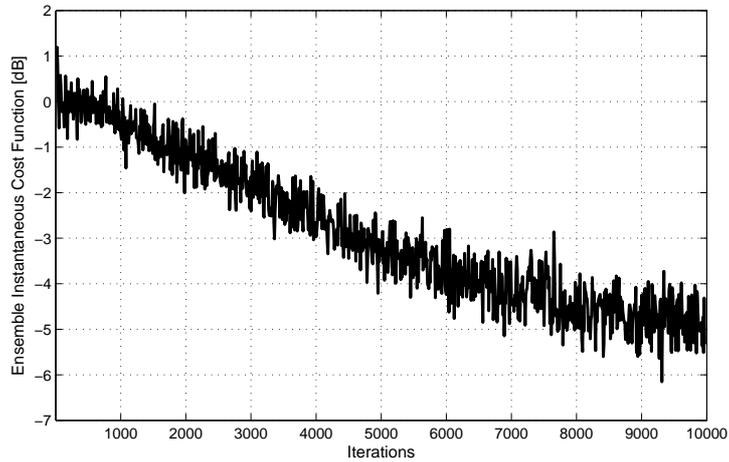}
	\caption{Learning curve of the quaternion-valued constant modulus algorithm.}
	\label{fig:learningGCMA}
\end{figure}

\begin{figure}[t]
	\centering
	\includegraphics[scale=0.6]{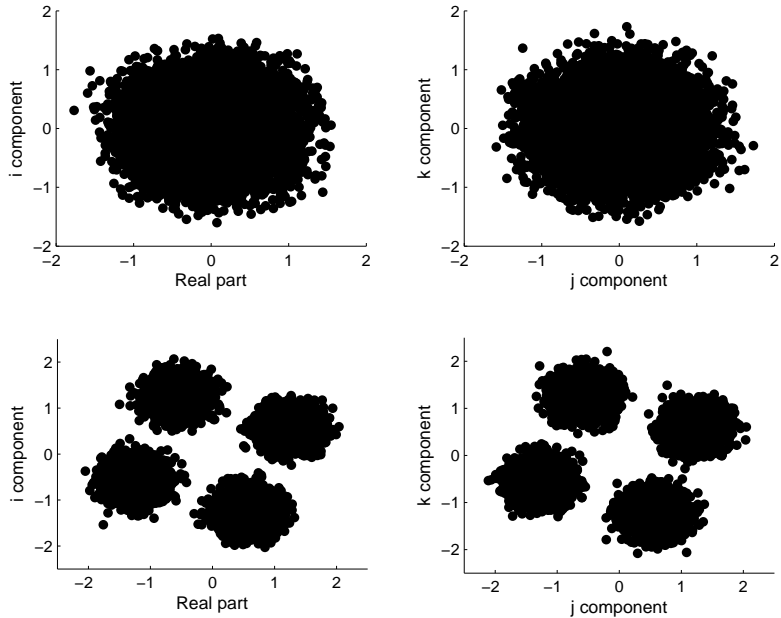}
	\caption{Scatter plots before and after equalization using the quaternion-valued constant modulus algorithm.}
	\label{fig:Qblind_equal}
\end{figure}

The learn curve in terms of the instantaneous cost function $\frac{1}{4}(|y[n]|^2-4)^2$ averaged over 200 runs is shown in Fig.~\ref{fig:learningGCMA}, where we can observe that the approximately constant modulus status has been reached. The scatter plots of constellation before and after equalization for the specific channel realization in (\ref{eqn:sim_fc}) are shown in Fig.~\ref{fig:Qblind_equal}, which clearly demonstrate that  the equalization operation has been successful. One note is the phase shift in the plot, which is normal as the blind equalization algorithm we employed here is ambiguous to an arbitrary phase shift.

%%%%%%%%%%%%%%%%%%%%%%%%%%%%%%%%%%%%%%%%%%%%%%%%%%%%%%%%%%%%%%%%%%%%%%%%%%%%%%%%%%%%%%%%%%%%%
\section{Conclusions}
\label{sec:concl}

A 4-D modulation scheme using quaternion-valued representation based on two antennas with different polarisation directions has been studied for wireless communications. Although quaternion-valued wireless communication systems have been investigated in the past and  research progress has been made in this promising area, a crucial missing link  is lack of effective and efficient quaternion-valued signal processing algorithms for channel equalization and interference suppression/beamforming. To fill this gap, two representative quaternion-valued signal processing algorithms have been introduced: one is the reference-signal based quaternion-valued least mean square algorithm with the associated Wiener solution, and other one is the blind quaternion-valued constant modulus algorithm. Three sets of simulation results were provided showing that both algorithms work effectively in the single-input-single-output and multiple-input-multiple-output cases and such a 4-D modulation scheme can therefore  be considered as a viable approach for future wireless communication systems.

%\bibliographystyle{C:/work/sheffield/IEEE}
%\bibliography{C:/work/sheffield/mybib}

\begin{thebibliography}{10}

\bibitem{isaeva95a}
O.~M. Isaeva and V.~A. Sarytchev,
\newblock ``Quaternion presentations polarization state,''
\newblock in {\em Proc. 2nd IEEE Topical Symposium of Combined
  Optical-Microwave Earth and Atmosphere Sensing}, Atlanta, US, April 1995, pp.
  195--196.

\bibitem{zetterberg77a}
L.~H. Zetterberg and H.~Br\"{a}ndstr\"{o}m,
\newblock ``Codes for combined phase and amplitude modulated signals in a
  four-dimensional space,''
\newblock {\em IEEE Transactions on Communications}, vol. COM-25, no. 29, pp.
  943--950, September 1977.

\bibitem{bihan04a}
N.~Le~Bihan and J.~Mars,
\newblock ``Singular value decomposition of quaternion matrices: a new tool for
  vector-sensor signal processing,''
\newblock {\em Signal Processing}, vol. 84, no. 7, pp. 1177--1199, 2004.

\bibitem{miron06a}
S.~Miron, N.~Le~Bihan, and J.~I. Mars,
\newblock ``Quaternion-{MUSIC} for vector-sensor array processing,''
\newblock {\em IEEE Transactions on Signal Processing}, vol. 54, no. 4, pp.
  1218--1229, April 2006.

\bibitem{gong11a}
X.~F. Gong, Z.~W. Liu, and Y.~G. Xu,
\newblock ``Direction finding via biquaternion matrix diagonalization with
  vector-sensors,''
\newblock {\em Signal Processing}, vol. 91, no. 4, pp. 821--831, 2011.

\bibitem{tao13a}
J.~W. Tao and W.~X. Chang,
\newblock ``A novel combined beamformer based on hypercomplex processes,''
\newblock {\em IEEE Transactions on Aerospace and Electronic Systems}, vol. 49,
  no. 2, pp. 1276--1289, 2013.

\bibitem{liu14e}
X.~R. Zhang, W.~Liu, Y.~G. Xu, and Z.~W. Liu,
\newblock ``Quaternion-valued robust adaptive beamformer for electromagnetic
  vector-sensor arrays with worst-case constraint,''
\newblock {\em Signal Processing}, vol. 104, pp. 274--283, November 2014.

\bibitem{liu14j}
M.~B. Hawes and W.~Liu,
\newblock ``Sparse vector sensor array design based on quaternionic
  formulations,''
\newblock in {\em Proc. of the European Signal Processing Conference}, Lisbon,
  Portugal, September 2014.

\bibitem{liu15a}
M.~B. Hawes and W.~Liu,
\newblock ``Design of fixed beamformers based on vector-sensor arrays,''
\newblock {\em International Journal of Antennas and Propagation}, 2015.

\bibitem{liu14f}
M.~D. Jiang, W.~Liu, and Y.~Li,
\newblock ``A general quaternion-valued gradient operator and its applications
  to computational fluid dynamics and adaptive beamforming,''
\newblock in {\em Proc. of the International Conference on Digital Signal
  Processing}, Hong Kong, August 2014.

\bibitem{liu14h}
M.~D. Jiang, W.~Liu, and Y.~Li,
\newblock ``A zero-attracting quaternion-valued least mean square algorithm for
  sparse system identification,''
\newblock in {\em Proc. of IEEE/IET International Symposium on Communication
  Systems, Networks and Digital Signal Processing}, Manchester, UK, July 2014.

\bibitem{liu14p}
M.~D. Jiang, Y.~Li, and W.~Liu,
\newblock ``Properties and applications of a restricted \mbox{HR} gradient
  operator,''
\newblock {\em arXiv:1407.5178 [math.OC]}, July 2014.

\bibitem{liu14n}
W.~Liu,
\newblock ``Antenna array signal processing for a quaternion-valued wireless
  communication system,''
\newblock in {\em Proc. the Benjamin Franklin Symposium on Microwave and
  Antenna Sub-systems (BenMAS)}, Philadelphia, US, September 2014.

\bibitem{wysocki09a}
B.~J. Wysocki and T.~A. Wysocki,
\newblock ``On an orthogonal space-time-polarization block code,''
\newblock {\em Journal of Communications}, vol. 4, no. 1, pp. 20--25, February
  2009.

\bibitem{compton81b}
R.~T. Compton,
\newblock ``The tripole antenna: An adaptive array with full polarization
  flexibility,''
\newblock {\em IEEE Transactions on Antennas and Propagation}, vol. 29, no. 6,
  pp. 944--952, November 1981.

\bibitem{nehorai99a}
A.~Nehorai, K.~C. Ho, and B.~T.~G. Tan,
\newblock ``Minimum-noise-variance beamformer with an electromagnetic vector
  sensor,''
\newblock {\em IEEE Transactions on Signal Processing}, vol. 47, no. 3, pp.
  601--618, March 1999.

\bibitem{zoltowski00a}
M.~D. Zoltowski and K.~T. Wong,
\newblock ``\mbox{ESPRIT}-based \mbox{2D} direction finding with a sparse
  uniform array of electromagnetic vector-sensors,''
\newblock {\em IEEE Transactions on Signal Processing}, vol. 48, pp.
  2205--2210, August 2000.

\bibitem{liu14l}
X.~R. Zhang, Z.~W. Liu, Y.~G. Xu, and W.~Liu,
\newblock ``Adaptive tensorial beamformer based on electromagnetic
  vector-sensor arrays with coherent interferences,''
\newblock {\em Multidimensional Systems and Signal Processing}, 2014, DOI:
  10.1007/s11045-014-0281-8.

\bibitem{hamilton66a}
W.~R. Hamilton,
\newblock {\em Elements of Quaternions},
\newblock Longmans, Green, \& co., 1866.

\bibitem{kantor89a}
I.~Kantor, A.~S. Solodovnikov, and A.~Shenitzer,
\newblock {\em Hypercomplex Numbers: an Elementary Introduction to Algebras},
\newblock Springer Verlag, New York, 1989.

\bibitem{ding01a}
Z.~Ding and Y.~Li,
\newblock {\em Blind Equalisation and Identification},
\newblock Signal Processing and Communications. CRC, New York, 2001.

\bibitem{vantrees02a}
H.~L. Van~Trees,
\newblock {\em Optimum Array Processing, Part IV of Detection, Estimation, and
  Modulation Theory},
\newblock Wiley, New York, 2002.

\bibitem{liu10g}
W.~Liu and S.~Weiss,
\newblock {\em Wideband Beamforming: Concepts and Techniques},
\newblock John Wiley \& Sons, Chichester, UK, 2010.

\bibitem{gooch86a}
R.~Gooch and J.~Lundell,
\newblock ``{CM array: an adaptive beamformer for constant modulus signals},''
\newblock in {\em Proc. IEEE International Conference on Acoustics, Speech, and
  Signal Processing}, New York, NY, 1986, pp. 2523--2526.

\bibitem{krim96}
H.~Krim and M.~Viberg,
\newblock ``Two decades of array signal processing research: the parametric
  approach,''
\newblock {\em IEEE Signal Processing Magazine}, vol. 13, no. 4, pp. 67--94,
  July 1996.

\bibitem{godara97b}
L.~C. Godara,
\newblock ``Application of antenna arrays to mobile communications, part ii:
  Beam-forming and direction-of-arrival estimation,''
\newblock {\em Proceedings of the IEEE}, vol. 85, no. 8, pp. 1195--1245, August
  1997.

\bibitem{johnson98a}
C.~R. Johnson, P.~Schniter, T.~J. Endres, J.~D. Behm, D.~R. Brown, and R.~A.
  Casas,
\newblock ``Blind equalization using the constant modulus criterion: A
  review,''
\newblock {\em Proceedings of the IEEE}, vol. 86, no. 10, pp. 1927--1950,
  October 1998.

\bibitem{chen07d}
S.~Chen, A.~Wolfgang, and L.~Hanzo,
\newblock ``Constant modulus algorithm aided soft decision directed scheme for
  blind space-time equalisation of \mbox{SIMO} channels,''
\newblock {\em Signal Processing}, vol. 87, pp. 2587--2599, November 2007.

\bibitem{liu10k}
L.~Zhang, W.~Liu, and R.~J. Langley,
\newblock ``A class of constant modulus algorithms for uniform linear arrays
  with a conjugate symmetric constraint,''
\newblock {\em Signal Processing}, vol. 90, pp. 2760--2765, September 2010.

\bibitem{haykin96a}
S.~Haykin,
\newblock {\em Adaptive Filter Theory},
\newblock Prentice Hall, Englewood Cliffs, New York, 3rd edition, 1996.

\end{thebibliography}

\end{document}